\documentclass{publ}

\usepackage[T1]{fontenc}
\usepackage{dfadobe}

\ifpdf \usepackage[pdftex]{graphicx} \pdfcompresslevel=9
\else \usepackage[dvips]{graphicx} \fi

\BibtexOrBiblatex
\electronicVersion
\PrintedOrElectronic

\usepackage[inkscapelatex=false]{svg}
\usepackage{booktabs}
\usepackage{amssymb}
\usepackage{pdfpages}
\usepackage{cite}
\usepackage{hyperref}

\title[{\textsc{PhiPlot}}]
{{\textsc{PhiPlot}}: A Web-Based Interactive EDA Environment for Atmospherically Relevant Molecules}

\author[M. Loukojärvi, A. Mahadevan, K. Haitsiukevich \& K. Puolamäki]
{\parbox{\textwidth}{\centering M.\,Loukojärvi$^{1}$\orcid{0009-0005-5719-0735}
		, A. Mahadevan$^{1}$\orcid{0000-0001-5401-5716} 
		, K. Haitsiukevich$^{1}$\orcid{0000-0001-8116-1292}
		and K. Puolamäki$^{1}$\orcid{0000-0003-1819-1047}
	}
	\\
	{\parbox{\textwidth}{\centering $^1$University of Helsinki, Helsinki, Finland\\
		}
	}
}

\begin{document}
	
	\teaser{
		\includegraphics[width=0.95\linewidth]{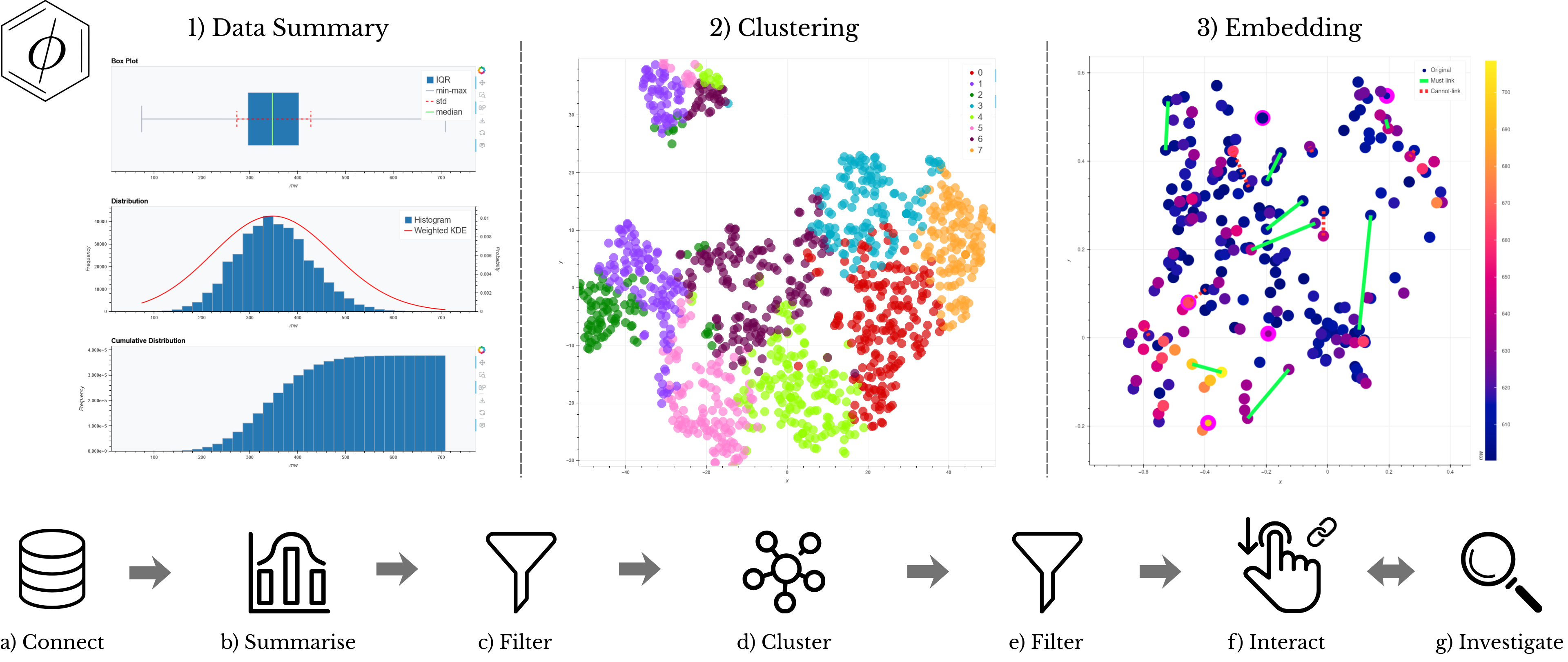}
		\centering
		\caption{Three main views of {\textsc{PhiPlot}} (1--3) and the corresponding Exploratory Data Analysis (EDA) pipeline (a--g). The application connects to a database of molecules, allows users to view summary statistics, filter, cluster, interactively embed and visualise the data.}
		\label{fig:teaser}
	}
	
	\maketitle
	
	\begin{abstract}
		Advances in computational chemistry have produced high-dimensional datasets on atmospherically relevant molecules. To aid exploration of such datasets, particularly for the study of atmospheric aerosol formation, we introduce {\textsc{PhiPlot}}: a web-based environment for interactive exploration and knowledge-based dimensionality reduction. The integration of visualisation, clustering, and domain knowledge-guided embedding refinement enables the discovery of patterns in the data and supports hypothesis generation. The application connects to an existing, evolving collection of molecular databases, offering an accessible interface for data-driven research in atmospheric chemistry. 



	\end{abstract}  
	
	\section{Introduction}
	
	In chemical sciences, advances in accessible frameworks and powerful hardware have made it possible to generate large, often high-dimensional, data sets of chemical structures and related properties via computational means \cite{ullah2024Molecular, chetry_molecules_2025}. This has created an increasing need to explore and understand these complex datasets \cite{orlov_high_2025, sandstrom_similarity-based_2025}. One such domain is the study of atmospheric aerosol formation. While recent works \cite{besel_atomic_2023, franzon_ether_2024, kahara_enhanced_2025} have produced extensive datasets for this domain, identifying the specific combinations of properties which trigger processes, such as particle formation in the atmosphere \cite{kerminen_atmospheric_2018}, remains an open problem. The study of this phenomenon is further constrained by the incompleteness of these datasets, as simulations with quantum-level accuracy for every potential molecule are computationally prohibitive \cite{kubecka_accurate_2024}.
	
	To bridge the gap between raw data and scientific insight, we present {\textsc{PhiPlot}}: a web-based Exploratory Data Analysis (EDA) environment designed to navigate these complex datasets. {\textsc{PhiPlot}} enables researchers to interactively explore atmospherically relevant chemical space by leveraging knowledge-based dimensionality reduction to support hypothesis generation, informed subsetting of molecules, and uncovering meaningful patterns within complex molecular datasets comprising molecules with quantitative properties.
	
	The application provides an easy-to-use, accessible interface. The user can get an overview of the data by accessing summary statistics of the available covariates, possibly with filters applied. The user can cluster the data and use the cluster labels as new features to further explore and filter the data. Finally, a subset of the data can be embedded in a two-dimensional plane with the ability to interactively constrain it. As shown by prior work by Chen, Gärtner and Paurat \cite{paurat_invis_2013,chen_scalable_2024}, interactivity is an important component for domain knowledge incorporation. Motivated by their work, our application allows the domain expert to interact with the data through domain-knowledge-guided interactive embedding constraints. The source code for {\textsc{PhiPlot}} is available at \url{https://github.com/edahelsinki/phiplot}.
	
	As a summary, our contributions are as follows:
	\textbf{(i)} We apply existing data analysis methods, such as knowledge-based dimensionality reduction, to atmospherically relevant molecules.
	\textbf{(ii)} We provide a web-based environment integrated with a curated collection of molecular databases.
	\textbf{(iii)} We conduct a small user study to verify the application's utility.

	\section{Data Analysis Methods}\label{sec:methods}
	
	{\textsc{PhiPlot}} incorporates molecular representations and fast methods for constrained dimensionality reduction discussed below.
	
	\paragraph*{Molecular Representations.}
	The data consists of molecules identified by SMILES strings encoding the 2D structures \cite{weininger_smiles_1988} and molecular properties obtained from experiments or simulations. The SMILES strings are converted into numerical feature vectors called molecular fingerprints. The fingerprinting process checks whether predefined substructures are present in a molecule or enumerates and hashes its fragments \cite{raghunathan_molecular_2022}. There are several general-purpose fingerprinting methods available through libraries like RDKit \cite{landrum_rdkit_2026} and novel techniques such as ATMOMACCS developed specifically for atmospheric molecules \cite{lind_interpretable_2026}.
	
	\paragraph*{Constrained Dimensionality Reduction.}
	The resulting molecular fingerprints are high-dimensional, sparse bit or integer vectors. To visualise them, we need to reduce their dimensionality by projecting them into a 2D space. This is what we mean by dimensionality reduction, and we refer to the resulting 2D representation as an embedding. Typical dimensionality reduction methods are unsupervised, as they require only feature vectors without targets \cite{ma2013Review} and provide only static embeddings. An extension to this is semi-supervised dimensionality reduction, where targets and further constraints are provided for a subset of the input data \cite{sheikhpour_survey_2017}. 
	
	Constrained Kernel PCA (cKPCA) \cite{oglic_interactive_2014, chen_scalable_2024} is a fast semi-supervised dimensionality reduction method that allows real-time interactivity. It enables the incorporation of domain knowledge about the molecules not captured by the initial embedding via interactive embedding constraints. The constraints are added by so-called control points and link constraints. Control points are points in the embedding for which the user provides a desired position. Additionally, there are two kinds of link constraints: must-links and cannot-links. Must-links assert that two points in the embedding should be closer to each other; conversely, cannot-links push two points further apart in the embedding space. 
	
	Another method offering knowledge-based dimensionality reduction is Least Square Projections (LSP) \cite{paulovich_least_2008}; however, it is slower and highly sensitive to local changes \cite{oglic_interactive_2014}.

	\section{Design and Implementation}
	
	\subsection{Requirements}\label{sec:requirements}
	
	Our objective is to accelerate data exploration and verification for both domain and data experts with an accessible, easy-to-use application. Our design goals are: 
	
	\noindent\textbf{DG1. Data accessibility.} Provide an interface to assess the data from several molecular databases in a single place; The user should not have to search for the relevant data from multiple sources or deal with cleaning the data for analysis. This accelerates the process of deriving meaningful insights from the data.\\
	\textbf{DG2. Application accessibility.} Implement a web-based framework for molecular data analysis; The user should be able to access the application without having to install or configure anything on their side. This facilitates the seamless adoption of the application into existing workflows.\\
	\textbf{DG3. Flexible data exploration.} Include a sufficient tool set for data exploration; The user should be able to toggle between granular unsupervised learning techniques like clustering and dimensionality reduction (focus) and more high-level summary statistics (context), promoting a holistic understanding of the data.\\
	\textbf{DG4. Interactive data analysis.} Make exploration functionality enhanced via constraint embeddings; The user should have the ability to include their domain knowledge by interacting with the visualisations produced by the dimensionality reduction. The human-in-the-loop nature ensures alignment between the algorithm's latent space and the domain expert's mental model.
	
	\subsection{The {\textsc{PhiPlot}} Architecture}\label{sec:architecture}
	
	{\textsc{PhiPlot}} is designed to allow users to summarise, filter, cluster, and interactively embed molecular datasets from an integrated data store of molecular databases. The application interface is composed of three main views as illustrated in Figure~\ref{fig:teaser}. Each view displays the data with progressively increasing levels of detail. Additionally, in each view, the user can apply domain-knowledge filters to subset the studied molecules. This workflow follows the well-established organisational principle of "overview, zoom, filter, details-on-demand" proposed by Shneiderman \cite{shneiderman_eyes_1996}. The navigation between the views follows the principles of Furnas \cite{furnas_effective_1997} and the application interface follows the core principles behind previous designs in visual analytics \cite{puolamaki_visually_2010, huang_va_2023}.
	
	\subsection{Technical Implementation}
	To support \textbf{DG1} the application provides direct access, without any user-side configuration, to the data store hosting the evolving collection of molecular databases. In the document-oriented MongoDB-backed NoSQL data store, each database is organised into one or more collections of molecules. Not enforcing a schema allows flexible storage of data like raw simulation outputs. {\textsc{PhiPlot}} interfaces with the database by fetching relevant fields for the selected documents and then internally transforms the data into tabular form. The data summary computations are handled by database-native aggregation pipelines, increasing the efficiency of the application interface by avoiding extra data transfer.
	
	The interface of the application is written in Python using Holoviz Panel and other HoloViz-maintained libraries \cite{rudiger_panel_2025}. The plotting back-end has been set up to use the Bokeh library \cite{bokeh}. To comply with \textbf{DG2}, the application is containerised with Docker and deployed with OpenShift as a web application. Our prototype implementation, including the database backend and visualisation tools, is hosted on the CSC --- IT Centre for Science server.
	
	\subsection{Application Views}\label{sec:views}
	
	Each view of the application has its own set of tools, as well as tools shared across views. The shared tools include the ability to fetch molecule documents from the databases, generate molecular fingerprints, apply filters to subset the molecules and search for molecules within the current view. Navigation between the views happens via tabs at the top menu bar.
	
	\paragraph*{Data Summary.} 
	The user can compute summary statistics for the available fields in the currently selected molecular collection. In addition to the computed numerical statistics, the interface provides a variety of visualisations. These include box plots, binned distributions and approximated KDE curves. For categorical fields, comparison box plots can be created for numerical comparison fields.
	
	\paragraph*{Clustering.} 
	The user can cluster the molecules based on their molecular fingerprints (see Section~\ref{sec:methods}). Here, the user has to fetch documents from the data store. After fetching molecules and generating fingerprints, the user can cluster a subset of thousands of molecules with a selection of standard clustering algorithms, like K-means and BIRCH. Additionally, to visualise the computed clusters, the application provides a selection of standard static embedding algorithms, such as PCA and t-SNE. To help evaluate clustering quality, the interface displays the Silhouette, Calinski-Harabasz, and Davis-Bouldin scores \cite{arbelaitz_extensive_2013}.
	
	\paragraph*{Interactive Embedding.}
	To be able to visually process the data, the available dataset is typically filtered and, if necessary, subsampled to a size in the hundreds, after which the molecular fingerprints are embedded. The interface provides access to the same static embedding methods as for clustering, as well as the interactive cKPCA and LSP embedding methods. The user can adjust the applicable hyperparameters for the embedding algorithms. In the case of the cKPCA method, this includes the selection of the kernel and the hyperparameters controlling it. To help evaluate how well the initial embedding preserves the local and global structure of the high-dimensional data, the interface displays results for trustworthiness, KNN preservation, Shepard correlation, and stress \cite{thrun_analyzing_2023}. The combination of the data summary, clustering and embedding views, and the ability to easily navigate between them supports \textbf{DG3}.
	
	To support \textbf{DG4}, the user can interactively create additional embedding constraints by using the cKPCA method with a suitable kernel.  Adding control points is achieved by selecting and dragging points in the embedding. As a point is dragged, the embedding updates in near real time. The update takes less than a second for $N\lesssim 500$ embedded points. Thus, it is possible to continuously see how moving one control point affects the positions of other points. Additionally, the strength at which the link constraint is enforced is a parameter the user can provide via a slider in the interface. Adjusting the strength of the link constraints updates the embedding similarly in near real-time.

	\section{Related Work}\label{sec:related_work}
	
	The current work is the most closely related to applications developed for domain-knowledge-guided visualisation.
	
	InVis and InVis 2.0 are examples of non-web-based applications using knowledge-based dimensionality reduction for general datasets \cite{paurat_invis_2013, chen_scalable_2024}. The applications support interactivity via control points and link-constraints. However, these applications require the user to load in their own data, they apply only dimensionality reduction, and do not provide built-in support for summary statistics or clustering algorithms. 
	
	Solvent Surfer adapts the InVis applications for solvent selection \cite{boobier_interactive_2025}. The Solvent Surfer app has a web-based interface and supports interactive features similar to the InVis applications. Solvent Surfer has access to an integrated solvent dataset and uses physical properties of the solvents as the embedding features. In contrast to our application, it lacks more comprehensive EDA and molecular representation functionality. Additionally, the deployed interface does not support real-time interactive embedding updates.
	
	$\chi$iplot is a web-based application for the analysis of chemical data  \cite{tanaka_chiiplot_2023}. It can compute static embeddings and produce comprehensive EDA plots, and it also supports clustering out of the box. The application has been deployed and is publicly available. However, $\chi$iplot does not support molecular fingerprinting nor interactive embedding methods. Moreover, it provides access to some example datasets lacking more comprehensive database integration.
	
	Table~\ref{tab:comparison} summarises available functionality in prior work and our application w.r.t. the design goal described in Section~\ref{sec:requirements}. To the best of our knowledge, no prior method explicitly addresses all of our design goals.
	Our application is unique in that it integrates with an existing database of atmospherically relevant molecules and can produce not only static but also interactive embeddings in addition to EDA features like summary statistics and clustering analysis in a publicly available web-based environment.
	
	\begin{table}[ht]
		\centering
		\caption{Comparison of PhiPlot with related work methods and tools against the design goals (DGs).}
		\label{tab:comparison}
		\begin{tabular}{lcccc}
			\toprule
			\textbf{Method / Tool} & \textbf{DG1} & \textbf{DG2} & \textbf{DG3} & \textbf{DG4} \\
			\midrule
			InVis / InVis 2.0 \cite{paurat_invis_2013, chen_scalable_2024} & & & & \checkmark \\
			Solvent Surfer \cite{boobier_interactive_2025} & \checkmark & \checkmark & & \checkmark \\
			$\chi$iplot \cite{tanaka_chiiplot_2023} & & \checkmark & \checkmark & \\
			\textbf{PhiPlot (Ours)} & \checkmark & \checkmark & \checkmark & \checkmark \\
			\bottomrule
		\end{tabular}
	\end{table}
	
	\newpage
	
	\section{User Study} 
	
	\begin{figure}
		\centering
		\includegraphics[width=\linewidth]{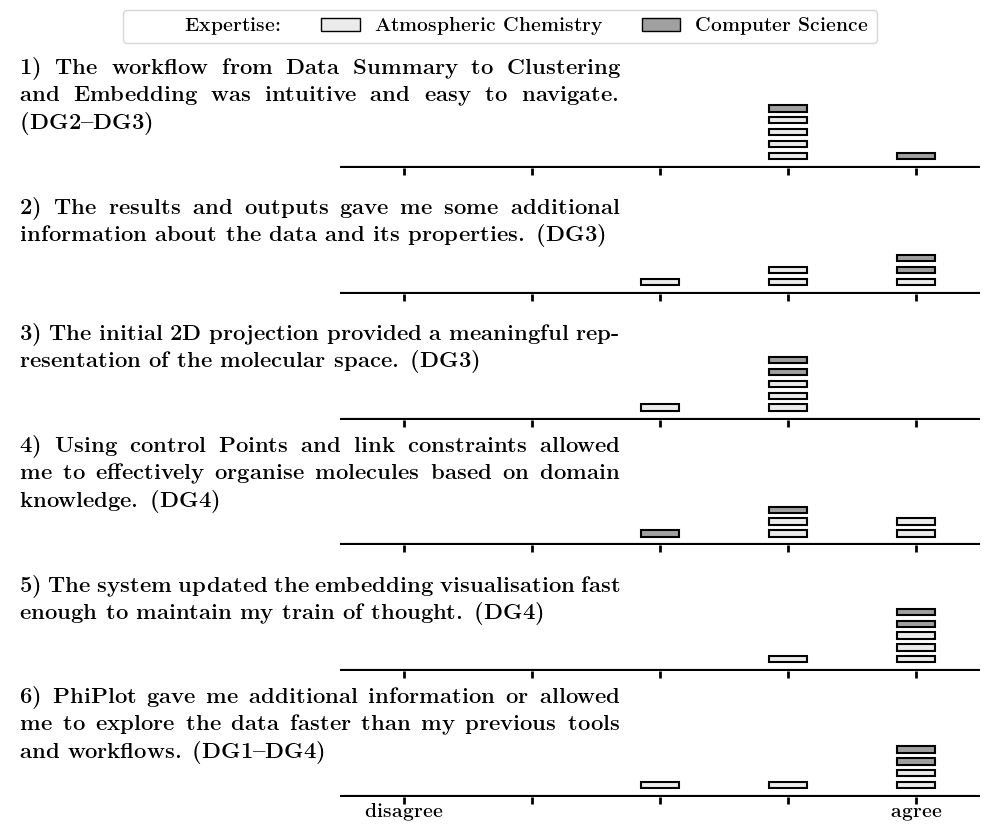}
		\caption{Results of the user survey on a 5-point Likert scale with results categorised by user expertise.}
		\label{fig:study}
	\end{figure}
	
	To evaluate how well {\textsc{PhiPlot}} meets the design goals set in Section \ref{sec:requirements} we conducted a small user study. The aim of the study was also to gain a better understanding of the utility, benefits, and limitations of {\textsc{PhiPlot}} for analysing molecular data and to identify ways in which the tool could be extended or adapted to fit more seamlessly into existing workflows. This was measured by asking users to rank six statements, shown in Figure \ref{fig:study}, about their subjective experience of using {\textsc{PhiPlot}} on a 5-point Likert scale, with 1 being disagree and 5 agree, and by asking them to answer two open questions about the benefits of {\textsc{PhiPlot}} compared to previous tools and workflows, and the most important improvements that could be made to {\textsc{PhiPlot}} to make it even more useful.
	
	The study was conducted either in person or remotely via Zoom, with a session lasting approximately 30 minutes. The study subjects were first given a brief introduction to the application, followed by a short tutorial on using the interface. The subjects were then given a set of tasks to complete based on the tutorial. The tasks were related to each of the views and the main features of the application described in Section~\ref{sec:views}. The interviewer guided the study subject in the right direction if they felt lost or did not understand something. After the task, the subjects were asked to fill out the survey. The study included a total of $N=6$ subjects. The subjects had expertise in either atmospheric chemistry (4 subjects; 2 senior researchers and 2 doctoral students) with varying levels of computer science expertise or computer science (2 subjects; doctoral students) with at least some familiarity with atmospheric chemistry. In the results, the categorisation is based on the main domain expertise. 
	
	The subjects largely agreed with the statements, validating our approach (see Figure \ref{fig:study}). The study highlights that interactions are required to make the initial static 2D projection more meaningful (statement 3). The answers demonstrate that domain experts, even if they were already familiar with the data, generally gained insights by using the application and leveraged interactions to embed domain knowledge (statements 4 and 6). Also, computer science experts with only basic familiarity with chemistry found the application insightful (statement 2) and helpful for fast data exploration (statements 1 and 5). All experts agreed that the implementation was fast (statement 5). 
	
	When asked about the pros of the app, the domain experts in atmospheric chemistry noted that {\textsc{PhiPlot}} was ``easy and intuitive to use'' and offered a ``fast way of looking at the general statistics and structure of the dataset.'' 
	The domain experts in computer science highlighted that the ``UI is fast and responsive,'' and there is a ``good selection of embeddings out-of-the-box.''
	When asked about areas for improvements, the chemistry experts would like {\textsc{PhiPlot}} to include ``[m]ore options for the datapoint search function'', and have the ability to ``select the group of molecules looked by [its] precursors''.
	The computer science experts would like the application to have the ability to ``[f]ilter and observe the number of certain functional groups.''

	\section{Conclusion and Future Work}
	
	This paper presents a working demo of {\textsc{PhiPlot}}, an application for exploring molecular data within a comprehensive EDA environment. {\textsc{PhiPlot}} offers low-barrier entry via its web interface and integration with a curated data store of molecular databases. The results of the user study confirm that our design goals were overall met and demonstrate the utility of {\textsc{PhiPlot}} in exploring atmospherically relevant molecules. Expert feedback indicates that the application is intuitive to use, provides meaningful insights into the data, and offers responsive interactivity, enabling the incorporation of domain knowledge. Experts agree that {\textsc{PhiPlot}} provides additional information about the data and enables faster exploration of the dataset than existing tools.
	
	The application functionality can be extended in several ways. Advanced filtering features based on chemical structure can simplify the interactions. Additionally, the application can be extended to explore and develop machine learning models trained to predict molecular properties.
	
	\textbf{Acknowledgements}
	Funded by the Research Council of Finland (decision 364226) and the Helsinki Institute for Information Technology HIIT. The authors thank CSC – IT Center for Science, Finland, and the Finnish Computing Competence Infrastructure (FCCI) for computational resources. We acknowledge research environment provided by ELLIS Institute Finland.
	
	\newpage
		
	\bibliographystyle{eg-alpha-doi} 
	\newcommand{\etalchar}[1]{$^{#1}$}

\end{document}